\documentclass{ws-rv9x6}
\usepackage{subfigure}   
\usepackage{ws-rv-thm}   
\usepackage{ws-rv-van}   
\makeindex

\def\simge{\mathrel{\rlap{\raise 0.511ex
       \hbox{$>$}}{\lower 0.511ex \hbox{$\sim$}}}}
\def\simle{\mathrel{\rlap{\raise 0.511ex
        \hbox{$<$}}{\lower 0.511ex \hbox{$\sim$}}}}

\begin{document}

\chapter[Thermal Effects in Dense Matter]{Thermal Effects in Dense Matter  \\ Beyond Mean Field Theory}\label{ra_ch1}

\author
{Constantinos Constantinou
}
\address{Institute for Advanced Simulation, Institute f\"ur Kernphysik, and J\"ulich Center for Hadron Physics, Forschungszentrum 
J\"ulich, D-52425 J\"ulich, Germany \\
c.constantinou@fz-juelich.de
}
\author[Constantinos Constantinou, Sudhanva Lalit and Madappa Prakash]{Sudhanva Lalit and Madappa Prakash}
\address{Department of Physics and Astronomy,\\
Ohio University, Athens, Ohio 45701, United States \\
sl897812@ohio.edu and
prakash@ohio.edu
}

\begin{abstract}
The formalism of next-to-leading order Fermi Liquid Theory is employed to calculate the thermal properties of symmetric nuclear and pure neutron matter in a relativistic many-body 
theory beyond the mean field level which includes two-loop effects. 
For all thermal variables, the semi-analytical next-to-leading order corrections reproduce results of the
exact numerical calculations for entropies per baryon up to 2. This corresponds to excellent agreement
down to subnuclear densities for temperatures up to $20$ MeV. 
In addition to providing physical insights, a rapid evaluation of the equation of state in the homogeneous 
phase of hot and dense matter is achieved through the use of the zero-temperature Landau effective mass function and its derivatives. 
\end{abstract}
\body

\tableofcontents

\section{Introduction}
\label{Intro_sec1}

Core-collapse supernovae, neutron stars from their birth to old age, and binary mergers involving neutron stars all pass through stages in which there are considerable variations in the 
baryon density, temperature, and lepton content.  Simulations of these astrophysical phenomena involve general relativistic hydrodynamics and neutrino transport with special relativistic 
effects. Convection, turbulence, magnetic fields, {\it etc}., also play crucial roles. 
The macroscopic evolution in each case is governed by microphysics involving strong, weak and electromagnetic interactions. Depending on the baryon density $n$, temperature $T$, 
and the lepton content of matter (characterized by $Y_{Le}=n_{Le}/n$ when neutrinos are trapped or by the net electron concentration $Y_e=n_e/n$ in neutrino-free matter), various 
phases of matter are encountered.  For sub-nuclear densities ($n\simle 0.1~{\rm fm}^{-3})$ and temperatures $T\simle 20~{\rm MeV}$, different inhomogeneous phases are encountered. 
A homogeneous phase of nucleonic and leptonic matter prevails at near- and supra-nuclear densities  ($n\simge 0.1~{\rm fm}^{-3})$ at all temperatures. With progressively increasing 
density, homogeneous matter may contain hyperons, quark matter and Bose condensates.  

Central to an understanding of the above astrophysical phenomena is the equation of state (EOS) of matter as a function of  $n,~T$, and $Y_{Le}$ (or $Y_e$) as it is as an integral part of  
hydrodynamical evolution, and controls electron capture and neutrino interactions in ambient matter. The EOS of dense matter has been investigated in the literature extensively, 
but for the most part those for use in the diverse physical conditions of relevance to astrophysical applications have been based on mean field theory in both non-relativistic potential or 
relativistic field-theoretical approaches.  A recent article honoring  Gerry Brown reviews the current status and advances made to date in the growing field of neutron star research
~[\refcite{latpra16}].  
 
The objective of this work is to assess the extent to which the model independent formalism of Fermi Liquid Theory (FLT)~[\refcite{baym91}] is able to accurately describe thermal 
effects in dense homogeneous nucleonic matter under degenerate conditions for models beyond mean field theory (MFT).  Recently, a next-to-leading order (NLO)  extension of the 
leading-order FLT was developed in Ref.~[\refcite{cons15}] incorporating its relativistic generalization in Ref.~[\refcite{baymchin76}]. The FLT+NLO formalism was applied to 
non-relativistic potential models with contact and finite-range interactions as well as  to relativistic models of dense matter at the mean field level  in Ref.~[\refcite{cons15}] .  Excellent 
agreement with the results of exact numerical calculations for all thermal variables was found with the semi-analytical FLT+NLO results. In this contribution, we present similar excellent 
agreement with the exact numerical results of a relativistic field-theoretical model beyond the MFT level that includes two-loop (exchange) effects recently reported in 
Ref.~[\refcite{zhang16}].  
The gratifying result is that the FLT+NLO formalism extends agreement with the exact numerical
results for all $n$ and $T$ for which the entropy per baryon $S \le 2$.  This means that, for $T \simle 20$ MeV,
 the method can adequately describe state variables down to a density of  $\sim 0.1~{\rm fm}^{-3}$. 
For densities below  $\sim 0.1~{\rm fm}^{-3}$, inhomogeneous phases occur for which a separate treatment is required. This development not 
only provides a check of time-consuming many-body calculations of dense matter  at finite temperature, but also serves to accurately (and, to rapidly) calculate thermal effects from a 
knowledge of the zero-temperature single-particle spectra for $S$ up to 2 for which effects of interactions are relatively important.   

The organization of this contribution is as follows. In Sec.~\ref{NLO}, we summarize the NLO formalism of FLT recently developed in Ref.~[\refcite{cons15}].  Section~\ref{BMFT} contains a brief description of the relativistic field-theoretical model that extends mean-field theory (MFT) to include two-loop  (TL) effects as implemented in Ref.~[\refcite{zhang16}].  Working formulas required for the evaluation of the degenerate-limit thermal effects (in particular, expressions for the single particle spectra) are  given in this section 
which also includes our results and associated discussion.   A summary of our work along with conclusions are in Sec.~\ref{SumCon}. Personal tributes to Gerry Brown from two of the authors (Constantinou and Prakash)  form the content of  Sec.~\ref{Tribs}.

\section{Next-to-Leading Order Fermi Liquid Theory}
\label{NLO}

The thermodynamics of fermion systems entails evaluation of integrals of the type
\begin{equation}
I = \int_0^{\infty}dp~ g(p)\frac{1}{1+\exp\left[\frac{\epsilon(p,n)-\mu}{T}\right]} \,,
\label{gofp}
\end{equation}
where $T$ is the temperature, $\mu$ is the chemical potential, and $\epsilon$ is the single-particle spectrum of the underlying model. The functional form of $g(p)$ is particular 
to the state property in question. Equivalently, we can write 
\begin{equation}
I = \int_0^{\infty}dy\frac{\phi(y)}{1+\exp(y-\eta)} \,,
\end{equation}
where 
\begin{eqnarray}
y &=& \frac{\epsilon(p,n)-\mathcal{U}(n)}{T}~~,~~\eta=\frac{\mu-\epsilon(p=0,n)}{T}  \\
\phi(y) &=& \frac{\mathcal{M}(p)Tg(p)}{p}~~,~~\mathcal{M}(p)=p\left(\frac{\partial \epsilon}{\partial p}\right)^{-1} \,.
\end{eqnarray}
Above, $\mathcal{U}(n)$ is inclusive of all those terms in the spectrum which depend only on the density $n$. The Landau effective mass function, $\mathcal{M}(p)$, and its derivatives with respect to momentum $p$  play crucial roles in determining the thermal effects. 

In the degenerate limit, characterized by large values of the parameter $\eta$, Sommerfeld's Lemma
\begin{equation}
I~ {\stackrel{\eta \gg 1}{\longrightarrow}} 
\int_0^{\eta}\phi(y)~dy  
    + \frac{\pi^2}{6}\left.\frac{d\phi}{dy}\right|_{y=\eta} 
   + \frac{7\pi^4}{360}\left.\frac{d^3\phi}{dy^3}\right|_{y=\eta} + \ldots    
\label{sommerfeld}
\end{equation}
can be used for the approximate evaluation of such integrals. Truncation of the series at the first term recovers results for cold matter; the second term produces the familiar 
Fermi Liquid Theory (FLT) corrections and the third term represents the next-to-leading order (NLO) extension to FLT. Owing to the asymptotic nature of the Sommerfeld formula, 
the expansion will, in general, diverge at higher orders unless all terms are retained.

The number density of single-species fermions with $\gamma$ internal degrees of freedom in 3 dimensions is  (throughout we use units in which $\hbar=1$)
\begin{equation}
n = \frac{\gamma}{2\pi^2}\int dp~\frac{p^2}{1+\exp\left(\frac{\epsilon-\mu}{T}\right)} \,.
\label{den-int}
\end{equation}
In the present context, we take $n$ to be an independent variable as is appropriate for a system that does not exchange particles with an external reservoir but whose total 
volume is allowed to change. Thus
\begin{equation}
n(T=0) = n(p_F) = \frac{\gamma p_F^3}{6\pi^2} = n(T) \simeq n(p_{\mu}) \,,
\label{den-equiv}
\end{equation}
where $p_F$ is the Fermi momentum and $n(p_{\mu})$ is the result of Eq. (\ref{den-int}) evaluated according to 
Eq.~(\ref{sommerfeld}). Perturbative inversion of Eq.~(\ref{den-equiv}) 
leads to 
\begin{equation}
p_{\mu} = p_F\left[1-\frac{\pi^2}{6}\frac{m^{*2}T^2}{p_F^4}\left(1+\frac{p_F}{m^*}\left.\frac{d\mathcal{M}}{dp}\right|_{p_F}\right)+\ldots \right] \,,
\label{pmu}
\end{equation}
where 
\begin{equation}
m^*=\mathcal{M}(p=p_F)
\label{mlandau}
\end{equation}
is the Landau effective mass.
The combination of Eq.~(\ref{pmu}) with the Sommerfeld expansion of the entropy density, formally given by
\begin{eqnarray}
s &=& \frac{\gamma}{2\pi^2}\int dp~p^2\left\{f(p)\ln f(p)-[1-f(p)]\ln [1-f(p)]\right\} \\
f(p) &=& \frac{1}{1+\exp\left[\frac{\epsilon(p)-\mu}{T}\right]}
\end{eqnarray}
yields an expression for $s$ in terms of quantities defined on the Fermi surface:
\begin{eqnarray}
s &=& \frac{\gamma p_Fm^*T}{6} - \frac{\gamma \pi^2}{15}\frac{m^{*3}T^3}{p_F^3}(1-L_F)   \\
   &=& 2anT - \frac{16}{5\pi^2}a^3nT^3(1-L_F) \,, \label{s-nlo}
\end{eqnarray}
where $a=\pi^2m^*/(2p_F^2)=\pi^2/(4T_F)$ is the level density parameter with $T_F$ denoting the Fermi temperature, and 
\begin{equation}
L_F \equiv \frac{7}{12}\left(\frac{p_F}{m^*}\left.\frac{\partial\mathcal{M}}{\partial p}\right|_{p_F}\right)^2
               +\frac{7}{12}\frac{p_F^2}{m^*}\left.\frac{\partial^2\mathcal{M}}{\partial p^2}\right|_{p_F}
               +\frac{3}{4}\frac{p_F}{m^*}\left.\frac{\partial\mathcal{M}}{\partial p}\right|_{p_F} \,.
\end{equation}
Then the entropy per particle is the simple ratio $S = s/n$ whereas the thermal energy, pressure and chemical potential are obtained via Maxwell's relations (the integrals below are performed at constant density):
\begin{eqnarray}
E_{th} &=& \int T~dS = aT^2- \frac{12}{5\pi^2}a^3T^4(1-L_F) 
\label{Eth-nlo} \\
P_{th} &=& -n^2\int \frac{dS}{dn}~dT \nonumber 
\label{Pth-nlo} \\
          &=& \frac{2}{3}anQT^2- \frac{8}{5\pi^2}a^3nQT^4\left(1-L_F+\frac{n}{2Q}\frac{dL_F}{dn}\right) \\
\mu_{th} &=& -\int \frac{ds}{dn}~dT   \nonumber  \\
               &=& -a\left(1-\frac{2Q}{3}\right)T^2+ \frac{4}{5\pi^2}a^3T^4\left[(1-L_F)(1-2Q)-n\frac{dL_F}{dn}\right] \,,
               \label{muth-nlo}
\end{eqnarray}
where 
\begin{equation}
Q = 1-\frac{3n}{2m^*}\frac{dm^*}{dn}.
\end{equation}

Other quantities of interest such as the specific heats at constant volume and pressure, and the thermal index are given by standard thermodynamics:
\begin{eqnarray}
C_V &=& T\left.\frac{dS}{dT}\right|_n = 2aT- \frac{48}{5\pi^2}a^3T^3(1-L_F)  \, 
\label{cv-nlo} \\
C_P &=& T\left.\frac{dS}{dT}\right|_P = 2aT+ \frac{T}{n^2} \frac{\left(\left.\frac{\partial P_{th}}{\partial T}\right|_n\right)^2}{\left.\frac{\partial P_{total}}{\partial n}\right|_T} \, 
\label{cp-nlo}\\
\Gamma_{th} &=& 1+\frac{P_{th}}{nE_{th}}=1+\frac{2Q}{3} - \frac{4}{5\pi^2}a^2nT^2\frac{dL_F}{dn} \,.
\label{gth-nlo}
\end{eqnarray}

Note that while the NLO terms in the  thermal quantities above have the same temperature dependences as those of a free Fermi gas, the accompanying density-dependent factors differ reflecting the effects of interactions.

\section{Application to Models Beyond Mean Field Theory}
\label{BMFT}

In this work, we investigate the degenerate-limit thermodynamics of a relativistic field-theoretical model in which the nucleon-nucleon (NN) interaction is mediated by the exchange of 
$\sigma$, $\omega$, $\rho$ and $\pi$ mesons (scalar, vector, iso-vector and pseudo-vector, respectively). Nonlinear self-couplings of the scalar field are also included. The model is described by the Lagrangian density~[\refcite{zhang16,walecka74,boguta77}]
\begin{eqnarray}
\mathcal{L} &=& \mathcal{L}_N + \mathcal{L}_{meson}  \\
\mathcal{L}_N &=& \bar N\left[i \gamma^{\mu}(\partial_{\mu} + i\frac{g_{\rho}}{2}\vec \rho_{\mu}\cdot \vec \tau+i g_{\omega} \omega_{\mu} \right. \nonumber \\
                        &-& \left. i \frac{g_A}{2f_{\pi}}\gamma_5\vec \tau \cdot \partial_{\mu}\vec \pi) - (M - g_{\sigma}\sigma)\right]N     \\
\mathcal{L}_{meson} &=& \frac{1}{2}\partial_{\mu}\sigma\partial^{\mu}\sigma-\left(\frac{1}{2}+\frac{\kappa_3}{3!}\frac{g_{\sigma}\sigma}{M}
                                      + \frac{\kappa_4}{4!}\frac{g_{\sigma}^2\sigma^2}{M^2}\right)m_{\sigma}^2\sigma^2  \nonumber \\
                                    &-& \frac{1}{4}V^{\mu\nu}V_{\mu\nu} + \frac{1}{2}m_{\omega}^2\omega_{\mu}\omega^{\mu}   \nonumber  \\ 
                                    &-& \frac{1}{4} B_{\mu\nu}B^{\mu\nu} + \frac{1}{2}m_{\rho}^2\vec \rho_{\mu}\cdot \vec \rho^{\mu}  \nonumber \\
                                    &+& \frac{1}{2} \partial^{\mu}\vec \pi \cdot \partial_{\mu}\vec \pi - \frac{1}{2}m_{\pi}^2\vec \pi \cdot \vec \pi \,,
\end{eqnarray}
where 
\begin{eqnarray}
V_{\mu\nu} &=& \partial_{\mu}\omega_{\nu} - \partial_{\nu}\omega_{\mu}   \\
B_{\mu\nu} &=& \vec \tau (\partial_{\mu}\vec \rho_{\nu}-\partial_{\nu}\vec \rho_{\mu})+ i \frac{g_{\rho}}{2}[\vec \rho_{\mu}\cdot \vec\tau,\vec \rho_{\nu}\cdot \vec \tau]
\end{eqnarray} 
are the field-strength tensors and $\tau$ are the $SU(2)$ isospin matrices. We use the masses $M=939$ MeV, $m_{\sigma}=550$ MeV, $m_{\omega}=783$ MeV, 
$m_{\rho}=770$ MeV and $m_{\pi}=138$ MeV, the couplings $g_{\sigma}=8.604$, $g_{\omega}=7.522$, $g_{\rho}=7.614$, $\kappa_3=4.84$ and $\kappa_4=-4.47$, 
the pion decay constant $f_{\pi}=93$ MeV and the nucleon axial current constant $g_A =1.26$ as in Ref.~[\refcite{zhang16}].

All thermodynamic quantities of interest can be derived from the grand potential density $\Omega$ which is  related to the pressure by $\Omega = -P$.  For an isotropic system in its rest-frame, the pressure is obtained from the diagonal elements of the spatial part of the energy-momentum tensor $T_{\mu\nu}=\partial \mathcal{L}/\partial(\partial_{\mu}\phi)~\partial_{\nu}\phi - g_{\mu\nu}\mathcal{L}$. 
In mean-field theory (MFT), the result is   
\begin{eqnarray}
P &=& \frac{1}{3}\langle T_{ii}\rangle  \nonumber \\
  &=& \frac{\gamma_{spin}}{3}\sum_i \int \frac{d^3p}{(2\pi)^3}\frac{p^2}{E^*(p)}f_i(p)
   + \frac{g_{\omega}^2}{2m_{\omega}^2}n^2  + \frac{g_{\rho}^2}{8m_{\rho}^2}(n_i-n_j)^2  \nonumber \\
  &-& \frac{m_{\sigma}^2}{g_{\sigma}^2}(M-M^*)\left[\frac{1}{2}+\frac{\kappa_3}{6M}(M-M^*)+\frac{\kappa_4}{24M^2}(M-M^*)^2\right] \,,
\end{eqnarray}
where
\begin{equation}
E^*(p) = (p^2+M^{*2})^{1/2} \,.
\end{equation}
In the mean-field approximation, the spectrum $\epsilon_i(p)$ that enters the Fermi distribution function $f_i(p)=\{1+\exp[(\epsilon_i(p)-\mu_i)/T]\}^{-1}$ is given by
\begin{equation}  
\epsilon_i(p) = \pm E^*(p) + \frac{g_{\omega}^2}{m_{\omega}^2}n + \frac{g_{\rho}^2}{4m_{\rho}^2}(n_i-n_j) \,.
\end{equation}
The +(-) sign corresponds to particles (antiparticles) and the subscripts $i,j$ to the two nucleon species. The Dirac effective mass $M^*$ results from the minimization 
of $\Omega$ with respect to the expectation value of the scalar field. 

The leading corrections to the mean-field $\Omega$ arise from two-loop (TL) exchanges of the mesons involved in the model. These corrections are given by (see, e.g.,~ [\refcite{chin77,zhang16}])
\begin{eqnarray}
\Omega_{ex,\sigma} &=& -\frac{\gamma_{spin}}{4} g_{\sigma}^{2} \int d\tau_p d\tau_q~ f_{s}(p,q) D(k;m_{\sigma}^*) \sum_i f_i(p)f_i(q)  \\
\Omega_{ex,\omega} &=& -\frac{\gamma_{spin}}{4} g_{\omega}^{2} \int d\tau_p d\tau_q~ f_{v}(p,q) D(k;m_{\omega}) \sum_i f_i(p)f_i(q)  \\
\Omega_{ex,\rho} &=& -\frac{\gamma_{spin}}{16} g_{\rho}^{2} \int d\tau_p d\tau_q~ f_{v}(p,q) D(k;m_{\rho})  \nonumber \\
                             &\times& \sum_i f_i(p)[f_i(q)+2f_j(q)]  \\
\Omega_{ex,\pi} &=& -\frac{\gamma_{spin}}{16} \left(\frac{g_{A}M^*}{f_{\pi}}\right)^2 \int d\tau_p d\tau_q~ f_{pv}(p,q) D(k;m_{\pi})  \nonumber \\
                           &\times& \sum_i f_i(p)[f_i(q)+2f_j(q)]  \,,
\end{eqnarray}
where 
\begin{eqnarray}
d\tau_p &=& \frac{d^3p}{(2\pi)^32E^*(p)} \,, \quad  
f_s(p,q) = 4(p^{\mu}q_{\mu}+M^{*2}) \,, \\
f_v(p,q) &=& 8(p^{\mu}q_{\mu}-2M^{*2})\,, \quad 
f_{pv}(p,q) = 16(p^{\mu}q_{\mu}-M^{*2})  \,,  \\
D(k;m) &=& \frac{1}{k^{\mu}k_{\mu}-m^2}~;~ k^{\mu}=p^{\mu}-q^{\mu}~,~p^{\mu}p_{\mu}=q^{\mu}q_{\mu}=M^{*2}.
\end{eqnarray}

The corresponding TL contributions to the single-particle spectrum [via $\epsilon_{ex}^i=\delta\Omega_{ex}/\delta n_i(p)$ with 
$\delta/\delta n_i(p)\int d^3p/(2\pi)^3f_i(p)=1$; $i$=nucleon species] are [\refcite{chin77,zhang16}] :
\begin{eqnarray}
\epsilon_{ex,\sigma}^i &=& -\frac{\gamma_{spin}}{8}\frac{g_{\sigma}^{2}}{E^*(p)} \int  d\tau_q~ f_{s}(p,q) D(k;m_{\sigma}^*) f_i(q)  \\
\epsilon_{ex,\omega}^i &=& -\frac{\gamma_{spin}}{8}\frac{g_{\omega}^{2}}{E^*(p)}\int d\tau_q~ f_{v}(p,q) D(k;m_{\omega}) f_i(q)  \\
\epsilon_{ex,\rho}^i &=& -\frac{\gamma_{spin}}{32}\frac{g_{\rho}^{2}}{E^*(p)} \int  d\tau_q~ f_{v}(p,q) D(k;m_{\rho}) [f_i(q)+2f_j(q)]  \\
\epsilon_{ex,\pi}^i &=& -\frac{\gamma_{spin}}{32E^*(p)} \left(\frac{g_{A}M^*}{f_{\pi}}\right)^2\int d\tau_q~ f_{pv}(p,q) D(k;m_{\pi}) [f_i(q)+2f_j(q)] \nonumber \\
\end{eqnarray}

Note that at the TL level, the self-interactions of the scalar field bestow upon it  an effective scalar-meson mass
\begin{equation}
m_{\sigma}^* = m_{\sigma}\left[1+\kappa_3\left(\frac{M-M^*}{M}\right)+\frac{\kappa_4}{2}\left(\frac{M-M^*}{M}\right)^2\right]^{1/2} \,,
\end{equation}
which is used in all exchange terms involving the $\sigma$- meson.  

\subsection{Two-loop calculations of dense nucleonic matter}

The degenerate limit formalism delineated in Sec.~\ref{NLO} requires for its implementation, in principle, only the $T=0$ parts of the spectrum [for $\mathcal{M}(p)$] and 
the pressure (for $C_P$ and $M^*$). Note, however, that for cold matter the statements $dP/d\sigma=0$ and $d\mathcal{E}/d\sigma=0$ are equivalent (being that at $T=0$, 
$P=-n~d\mathcal{E}/dn$) and that the energy density $\mathcal{E}$ is somewhat easier to minimize with respect to $\sigma$ in order to obtain $M^*$.  We therefore opt to work with the latter. Confining ourselves to symmetric nuclear matter (SNM) 
and pure neutron matter (PNM) in the interest of simplicity, we have for the energy density (in the notation of Ref.~[\refcite{chin77}])
\begin{eqnarray}
\mathcal{E}_{TL} &=& \mathcal{E}_{MFT} +\mathcal{E}_{ex,\sigma}+\mathcal{E}_{ex,\omega}+\mathcal{E}_{ex,\rho}+\mathcal{E}_{ex,\pi}  \label {eden}\\
\mathcal{E}_{MFT} &=& 2\gamma_{iso}\int_0^{p_F}\frac{d^3p}{(2\pi)^3}E^* +\frac{1}{2}\frac{g_{\omega}^2}{m_{\omega}^2}n^2
                 +\frac{(1-\gamma_{charge})}{8}\frac{g_{\rho}^2}{m_{\rho}^2}n^2  \nonumber  \\
               &+& \frac{m_{\sigma}^2}{g_{\sigma}^2}(M-M^*)^2\left[\frac{1}{2}+\frac{\kappa_3}{3!}\left(\frac{M-M^*}{M}\right)
                +\frac{\kappa_4}{4!}\left(\frac{M-M^*}{M}\right)^2\right]  \\
\mathcal{E}_{ex,\sigma} &=& \gamma_{iso}\frac{g_{\sigma}^2}{(2\pi)^4}M^{*4}\left[\frac{1}{4}(x\eta-\ln\xi)^2+\left(1-\frac{w_{\sigma}^*}{4}\right)I(w_{\sigma}^*)\right]   \\
\mathcal{E}_{ex,\omega} &=& \gamma_{iso}\frac{g_{\omega}^2}{(2\pi)^4}M^{*4}\left[\frac{1}{2}(x\eta-\ln\xi)^2-\left(1+\frac{w_{\omega}}{2}\right)I(w_{\omega})\right]   \\
\mathcal{E}_{ex,\rho} &=& \frac{(\gamma_{iso}+4\gamma_{charge})}{4}\frac{g_{\rho}^2}{(2\pi)^4}M^{*4}\left[\frac{1}{2}(x\eta-\ln\xi)^2-\left(1+\frac{w_{\rho}}{2}\right)I(w_{\rho})\right] 
\nonumber  \\ \\
\mathcal{E}_{ex,\pi} &=& \frac{(\gamma_{iso}+4\gamma_{charge})}{4} \left(\frac{g_AM^*}{f_{\pi}}\right)^2\frac{M^{*4}}{(2\pi)^4}\left[(x\eta-\ln\xi)^2-w_{\pi}I(w_{\pi})\right]  \nonumber \\
\end{eqnarray}
and for the spectrum [via $\epsilon_{ex,i}=\delta\mathcal{E}_{ex,i}/\delta n$; $i$=meson] 
\begin{eqnarray}
\epsilon_{TL} &=& \epsilon_{MFT} +\epsilon_{ex,\sigma}+\epsilon_{ex,\omega}+\epsilon_{ex,\rho}+\epsilon_{ex,\pi} 
                 \label{spectra} \\
\epsilon_{MFT} &=&  M^*e +\frac{g_{\omega}^2}{m_{\omega}^2}n +\frac{(1-\gamma_{charge})}{4}\frac{g_{\rho}^2}{m_{\rho}^2}n  \\
\epsilon_{ex,\sigma} &=& \frac{g_{\sigma}^2}{(2\pi)^2}\frac{M^*}{2e}\left[\frac{1}{2}(x\eta-\ln\xi)+\left(1-\frac{w_{\sigma}^*}{4}\right)\frac{2}{r}J(w_{\sigma}^*)\right]  \\
\epsilon_{ex,\omega} &=& \frac{g_{\omega}^2}{(2\pi)^2}\frac{M^*}{2e}\left[(x\eta-\ln\xi)-\left(1+\frac{w_{\omega}}{2}\right)\frac{2}{r}J(w_{\omega})\right]  \\
\epsilon_{ex,\rho} &=& \frac{(\gamma_{iso}+\gamma_{charge})}{4}\frac{g_{\rho}^2}{(2\pi)^2}\frac{M^*}{2e}\left[(x\eta-\ln\xi)
                   -\left(1+\frac{w_{\rho}}{2}\right)\frac{2}{r}J(w_{\rho})\right]    \\
\epsilon_{ex,\pi} &=& \frac{(\gamma_{iso}+\gamma_{charge})}{4} \left(\frac{g_AM^*}{f_{\pi}}\right)^2\frac{M^*}{2e(2\pi)^2}
                 \left[2(x\eta-\ln\xi)-w_{\pi}\frac{2}{r}J(w_{\pi})\right]  \,, \nonumber \\
\end{eqnarray}
where 
\begin{eqnarray}
\gamma_{iso} &=& \left\{\begin{array}{ll}
                        1 & \mbox{PNM}  \\
                        2 & \mbox{SNM}
                        \end{array}\right.  ~~,~~
\gamma_{charge} = \left\{\begin{array}{ll}
                        0 & \mbox{PNM}  \\
                        1 & \mbox{SNM}
                        \end{array}\right.      \\   
x &=& \frac{p_F}{M^*}  ~~,~~\eta = (1+x^2)^{1/2}  ~~,~~\xi =  x+\eta    \\
r &=& \frac{p}{M^*}  ~~,~~e = (1+r^2)^{1/2}  ~~,~~t =  r+e     \\
w_i &=& \frac{m_i^2}{M^{*2}} ~~;~~ i=\sigma^*,\omega,\rho,\pi    \\
I(w_i) &=& \int_1^{\xi}\int_1^{\xi}dz~dy\left(1-\frac{1}{z^2}\right)\left(1-\frac{1}{y^2}\right)\ln\left[\frac{(zy-1)^2+w_izy}{(z-y)^2+w_izy}\right]  \\
J(w_i) &=& \int_1^{\xi}dz\left(1-\frac{1}{z^2}\right)\ln\left[\frac{(zt-1)^2+w_izy}{(z-t)^2+w_izy}\right]  .
\end{eqnarray}
With the inclusion of the TL contributions and noting that $M^*=M-g_{\sigma}\sigma$, the Dirac effective mass $M^*$ is determined by solving 
\begin{eqnarray}
\partial\mathcal{E}_{TL}/\partial M^* &=& 0 \quad {\rm with} \label{M*Dirac} \\ 
\frac{\partial\mathcal{E}_{MFT}}{\partial M^*} &=& 2\gamma_{iso}\int_0^{p_F}\frac{d^3p}{(2\pi)^3}\frac{M^*}{E^*}-\frac{m_{\sigma}^2}{g_{\sigma}^2}(M-M^*)
\nonumber \\
     &\times& \left[1+\frac{\kappa_3}{2}\left(\frac{M-M^*}{M}\right)+\frac{\kappa_4}{6}\left(\frac{M-M^*}{M}\right)^2\right] \\
\frac{\partial\mathcal{E}_{ex,\sigma}}{\partial M^*} &=& \frac{4\mathcal{E}_{ex,\sigma}}{M^*}+\gamma_{iso}\frac{g_{\sigma}^2}{(2\pi)^4}M^{*4} \nonumber \\
&\times&\left[-\frac{x^3}{M^*\eta}(x\eta-\ln\xi)+\left(1-\frac{w_{\sigma}^*}{4}\right)\frac{\partial I_{\sigma^*}}{\partial M^*}+\frac{w_{\sigma}^*}{2M^*}I_{\sigma^*}\right] \\
\frac{\partial\mathcal{E}_{ex,\omega}}{\partial M^*} &=& \frac{4\mathcal{E}_{ex,\omega}}{M^*}+\gamma_{iso}\frac{g_{\omega}^2}{(2\pi)^4}M^{*4} \nonumber \\
     &\times& \left[-\frac{2x^3}{M^*\eta}(x\eta-\ln\xi)-\left(1+\frac{w_{\omega}}{2}\right)\frac{\partial I_{\omega}}{\partial M^*}+\frac{w_{\omega}}{M^*}I_{\omega}\right] \\
\frac{\partial\mathcal{E}_{ex,\rho}}{\partial M^*} &=& \frac{4\mathcal{E}_{ex,\rho}}{M^*}+\frac{(\gamma_{iso}+4\gamma_{charge})}{4}\frac{g_{\rho}^2}{(2\pi)^4}M^{*4} \nonumber \\
     &\times& \left[-\frac{2x^3}{M^*\eta}(x\eta-\ln\xi)-\left(1+\frac{w_{\rho}}{2}\right)\frac{\partial I_{\rho}}{\partial M^*}+\frac{w_{\rho}}{M^*}I_{\rho}\right] \\
\frac{\partial\mathcal{E}_{ex,\pi}}{\partial M^*} &=& 
            \frac{6\mathcal{E}_{ex,\pi}}{M^*}+\frac{(\gamma_{iso}+4\gamma_{charge})}{4}\left(\frac{g_AM^*}{f_{\pi}}\right)^2\frac{M^{*4}}{(2\pi)^4} \nonumber \\
     &\times& \left[-\frac{4x^3}{M^*\eta}(x\eta-\ln\xi)-w_{\pi}\frac{\partial I_{\pi}}{\partial M^*}+\frac{2w_{\pi}}{M^*}I_{\pi}\right] \,,
\end{eqnarray}
where
\begin{eqnarray}
\frac{\partial I(w_{\sigma^*})}{\partial M^*} &=& \frac{1}{2M^*}\left\{w_{\sigma^*}\left(1-\frac{M^*}{m_{\sigma}^*}\frac{dm_{\sigma}^*}{dM^*}\right)\right.  \nonumber \\
             &\times& \int_1^{\xi}\int_1^{\xi}dy~dz\frac{1}{yz}\frac{(1-z^2)^2(1-y^2)^2}{[(zy-1)^2+w_{\sigma^*}zy][(z-y)^2+w_{\sigma^*}zy]}  \nonumber \\
              &-& \left. \frac{x\xi}{\eta}\left(1-\frac {1}{\xi^2}\right)J(w_{\sigma^*};t\rightarrow \xi) \right\} \label{disigdm} \\
\frac{dm_{\sigma}^*}{dM^*} &=& -\frac{m_{\sigma}^2}{2Mm_{\sigma}^*}\left[\kappa_3+\kappa_4\left(\frac{M-M^*}{M}\right)\right]  \\
\frac{\partial I(w_i)}{\partial M^*} &=& \frac{1}{2M^*}\left\{w_i\int_1^{\xi}\int_1^{\xi}dy~dz\frac{1}{yz}
                          \frac{(1-z^2)^2(1-y^2)^2}{[(zy-1)^2+w_izy][(z-y)^2+w_izy]} \right. \nonumber \\
              &-& \left. \frac{x\xi}{\eta}\left(1-\frac{1}{\xi^2}\right)J(w_i;t\rightarrow \xi) \right\} \label{didm}  ~~;~~ i=\omega,\rho,\pi .
\end{eqnarray}
To obtain Eqs. (\ref{disigdm}) and (\ref{didm}), the 2-dimensional Leibniz rule 
\begin{eqnarray}
\frac{d}{dt}\int_{x_0(t)}^{x_1(t)}\int_{y_0(t)}^{y_1(t)}F(x,y,t)dx~dy &=& 
             \int_{y_0}^{y_1}\left[F(x_1)\frac{\partial x_1}{\partial t}-F(x_0)\frac{\partial x_0}{\partial t}\right]dy  \nonumber \\
            &+& \int_{x_0}^{x_1}\left[F(y_1)\frac{\partial y_1}{\partial t}-F(y_0)\frac{\partial y_0}{\partial t}\right]dx  \nonumber \\
            &+& \int_{x_0}^{x_1}\int_{y_0}^{y_1}\frac{\partial F}{\partial t}dx~dy  
\end{eqnarray}
was employed.

\subsection*{Numerical notes}

The various integrals above are readily calculated by using the Gauss-Legendre quadrature method~[\refcite{AS72}]. The results reported below were calculated using 32 points and weights in each dimension although use of 16 points and weights was found to be adequate.   The derivatives of ${\mathcal M}(p)$ and $L_F(n)$ were calculated using the 5-point rule~[\refcite{AS72}].  Root finding was accomplished by the Newton-Raphson scheme. All numerical results of our Fortran code were also verified  by using Mathematica. 

\subsection{Results and discussion}
%

\begin{figure}[htb]
\centerline{\includegraphics[width=11cm]{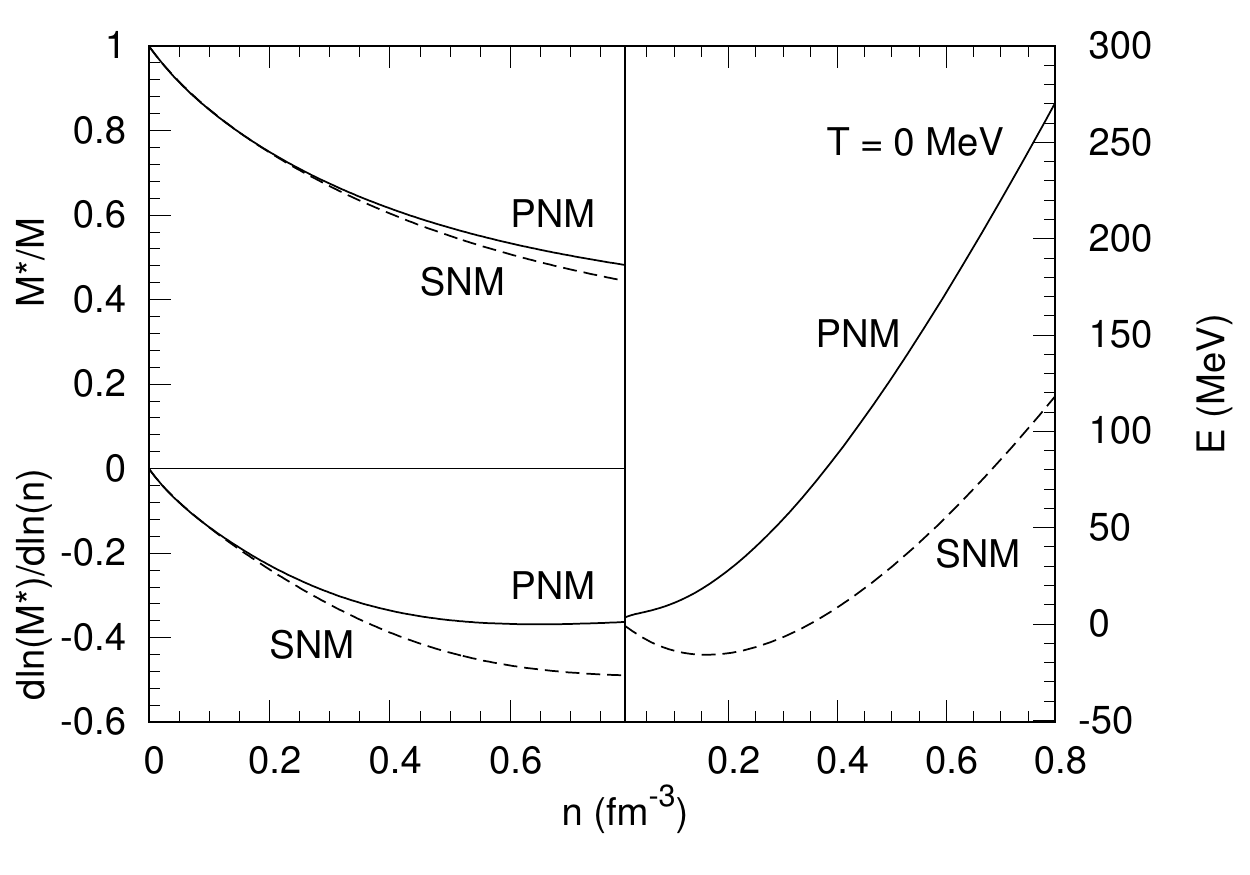}}
\vspace*{-0.25in}
\caption{Upper left panel: Dirac effective masses $M^*$ [Eq.~(\ref{M*Dirac})] scaled with the vacuum nucleon mass vs density $n$ in symmetric nuclear matter (SNM) and  pure neutron matter (PNM) 
at temperature $T=0$.
Lower left  panel: Logarithmic derivatives of $M^*$ w.r.t $n$. Right panel: Energy per particle $E=\partial {\mathcal E}/\partial n - M$ vs  $n$ in SNM and PNM at $T=0$.}
\label{msom_E}
\end{figure}

The variational procedure $\partial {\mathcal E}_{TL}/\partial M^* = 0$ in Eq.~(\ref{M*Dirac}) minimizes the energy density of the system and results in the optimal  baryon (Dirac) effective mass $M^*$ at each  $n$.  
(This minimization condition also yields the expectation value of the scalar field $\sigma=(M-M^*)/g_\sigma$.) The  values of $M^*$ in SNM and PNM are shown in the upper 
left panel of Fig.~\ref{msom_E}. Use of these $M^*$'s in the expressions for the energy density in Eq.~(\ref{eden}) allows for a calculation of the energy per particle, and are shown in the right 
panel of Fig.~\ref{msom_E} for SNM and PNM at $T=0$.  These results yield good agreement with nuclear and neutron star phenomenology~[\refcite{zhang16}]. 
The TL contributions to the energy density play a significant role in determining $M^*(n)$. The pattern $M^*({\rm PNM}) \geq M^*({\rm SNM})$ for a given baryon density stems from 
the isospin-invariant nucleon-nucleon interactions employed in the model. An MFT calculation - that is, without the TL terms in Eq.~(\ref{eden}) - that yields closely resembling $E$ vs. $n$ curves 
shown here through a readjustment of the various coupling strengths produces $M^*$ curves that vary more steeply with density (not shown here, but see Ref.~[\refcite{zhang16}]).   
As $M^*(n)$ and its logarithmic derivative w.r.t. $n$ (lower left panel of Fig.~\ref{msom_E}) enter prominently in determining the thermal properties, contrasts between different levels 
of theoretical approximations (MFT vs MFT+TL in our case here) are afforded. 

\begin{figure}[htb]
\centerline{\includegraphics[width=11cm]{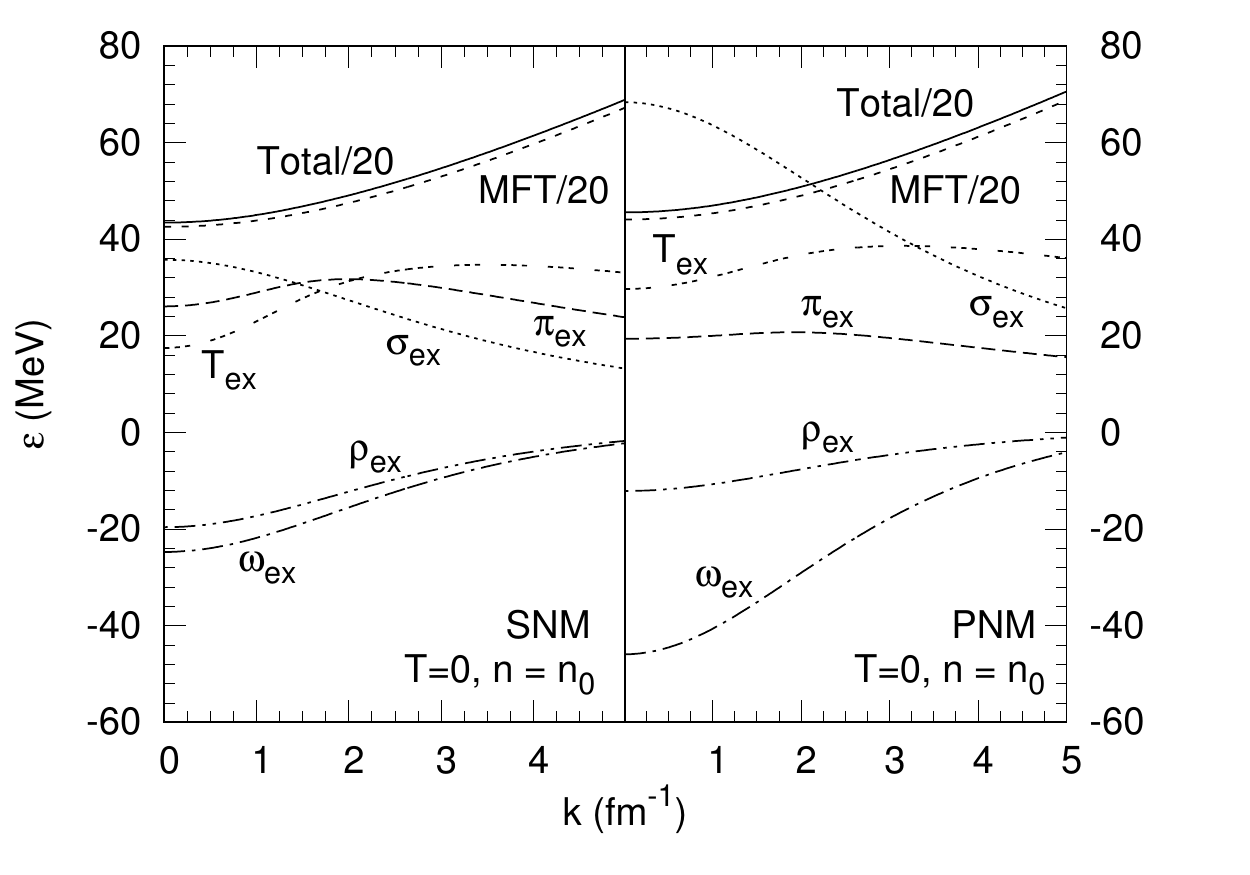}}
\vspace*{-0.25in}
\caption{
Contributions from MFT and TL terms [from Eq.~(\ref{spectra})] involving the exchange of $\sigma$, $\omega$, $\rho$, and $\pi$ mesons to the total 
single-particle energy vs. wave number in SNM and PNM at the baryon density $n=n_0=0.16~{\rm fm}^{-3}$.}
\label{spectra_n0}
\end{figure}

Under degenerate conditions for which $T/T_F \ll 1$, thermal effects depend sensitively on  details of the single-particle spectrum near the Fermi surface. The various contributions to the $T=0$ single-particle spectra in SNM and PNM 
are shown in Fig.~\ref{spectra_n0} at $n=n_0=0.16~{\rm fm}^{-3}$. Note that the dominant contribution from the MFT part in Eq.~(\ref{spectra}) has been divided by a factor of 20 to fit within 
the figure where contributions from the exchange of $\sigma$, $\omega$, $\rho$, and $\pi$ mesons from Eq.~(\ref{spectra}) are also shown. Although subdominant in their contributions to the 
spectra, the exchange contributions significantly alter the $M^*(n)$ curves from those of MFT and hence the MFT term of $\epsilon(p)$.  

\begin{figure}[htb]
\centerline{\includegraphics[width=11cm]{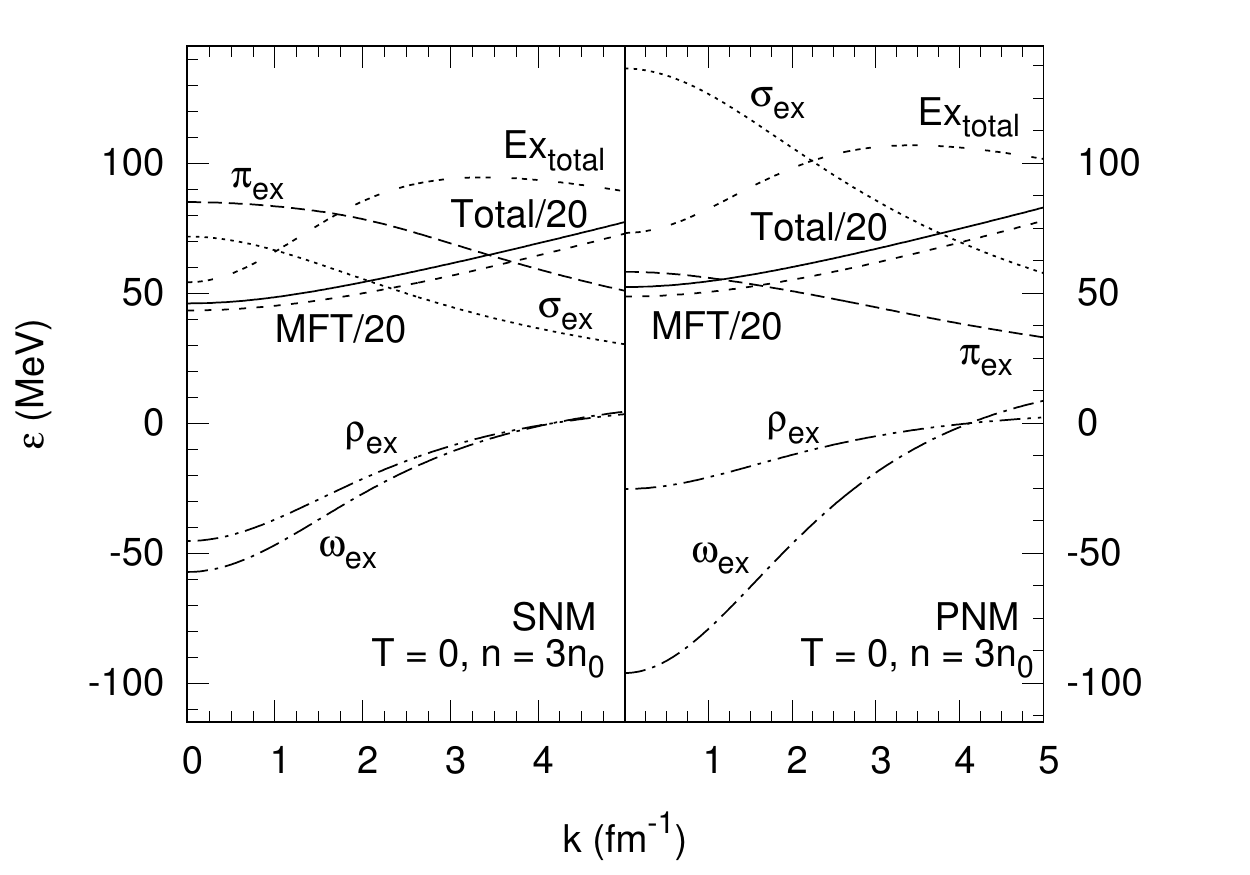}}
\vspace*{-0.15in}
\caption{Same as Fig.~\ref{spectra_n0}, but for $n=3n_0$.}
\label{spectra_3n0}
\end{figure}

Depending on the density, the magnitude and slope of the single-particle spectrum are also altered from its MFT contribution as can be seen in Fig.~\ref{spectra_3n0} where results 
for $n=3n_0$ are shown.   Such differences will be reflected in the thermal properties, particularly in the NLO terms of FLT.  

\begin{figure}[htb]
\centerline{\includegraphics[width=10cm]{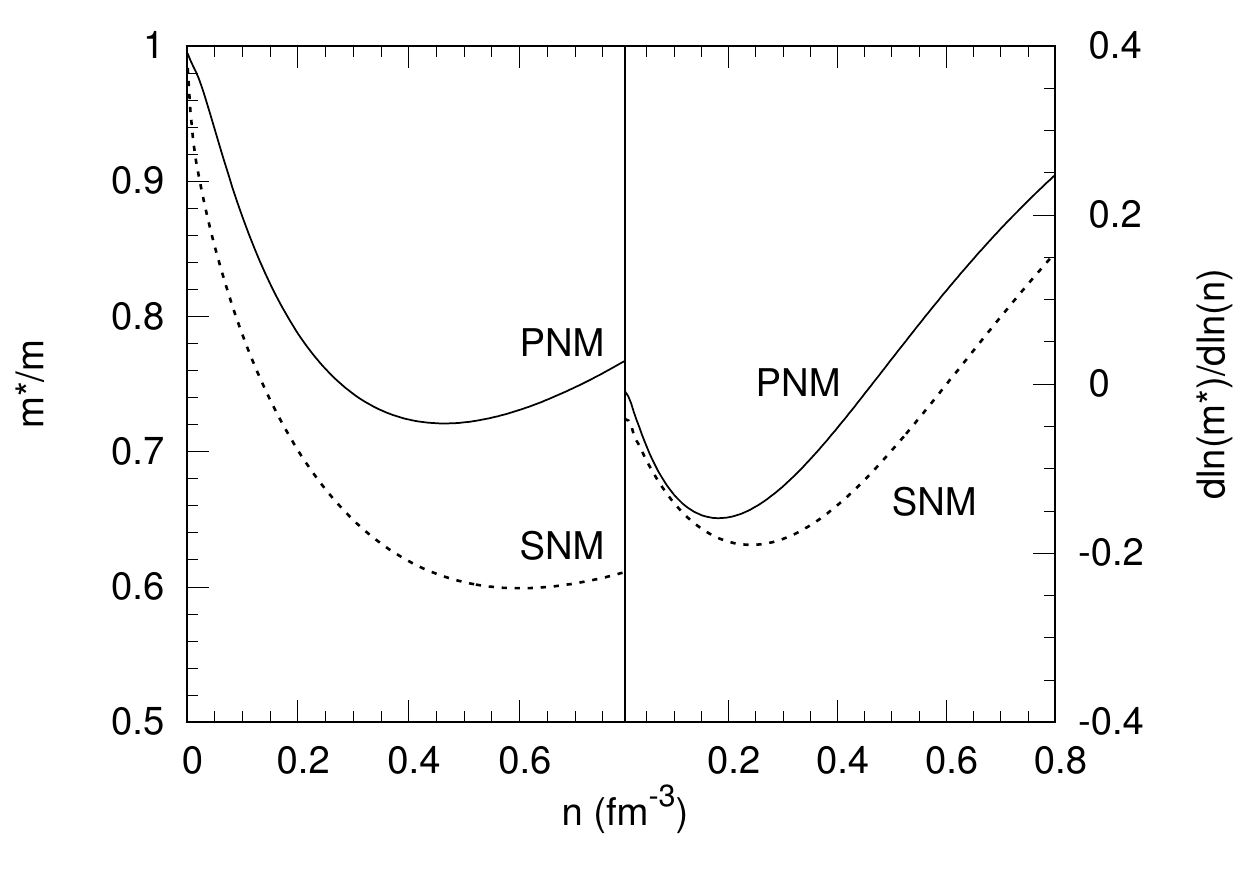}}
\vspace*{-0.2in}
\caption{Left panel: Landau effective masses from Eq.~(\ref{mlandau}) scaled with the vacuum nucleon mass vs density in SNM and PNM.
Right panel: Logarithmic derivatives of the Landau effective masses w.r.t. density.}
\label{msom}
\end{figure}

The Landau effective masses $m^*(n)$ from Eq.~(\ref{mlandau})  scaled with the vacuum nucleon mass are shown in the left panel of Fig.~\ref{msom} as functions of density in SNM and PNM. 
The associated logarithmic derivatives are in the right panel of this figure. The TL results are substantially larger than those of MFT for the same $n$ (see Ref.~[\refcite{zhang16}]). 
The non-monotonic behaviors and the minima in the $m^*(n)$ curves are characteristic of relativistic field theoretical models in which $M^*(n)$ continually decreases with increasing $n$. 
Together with the derivatives of the Landau effective mass function required at NLO in FLT, $m^*(n)$ and its logarithmic derivative play important roles in improving  the accuracy of the 
degenerate limit thermodynamics.

We turn now to compare the thermal properties from FLT and FLT+NLO with those from the exact numerical results of Ref.~[\refcite{zhang16}]  for the TL calculations at temperatures 
of $T=20$ and 50 MeV, respectively. In all cases, comparisons shown for $T=50$ MeV highlight the onset of semi- or non-degenerate regions in density for which results of degenerate 
limit FLT and FLT+NLO begin to become inadequate.

\begin{figure}[htb]
\centerline{\includegraphics[width=11cm]{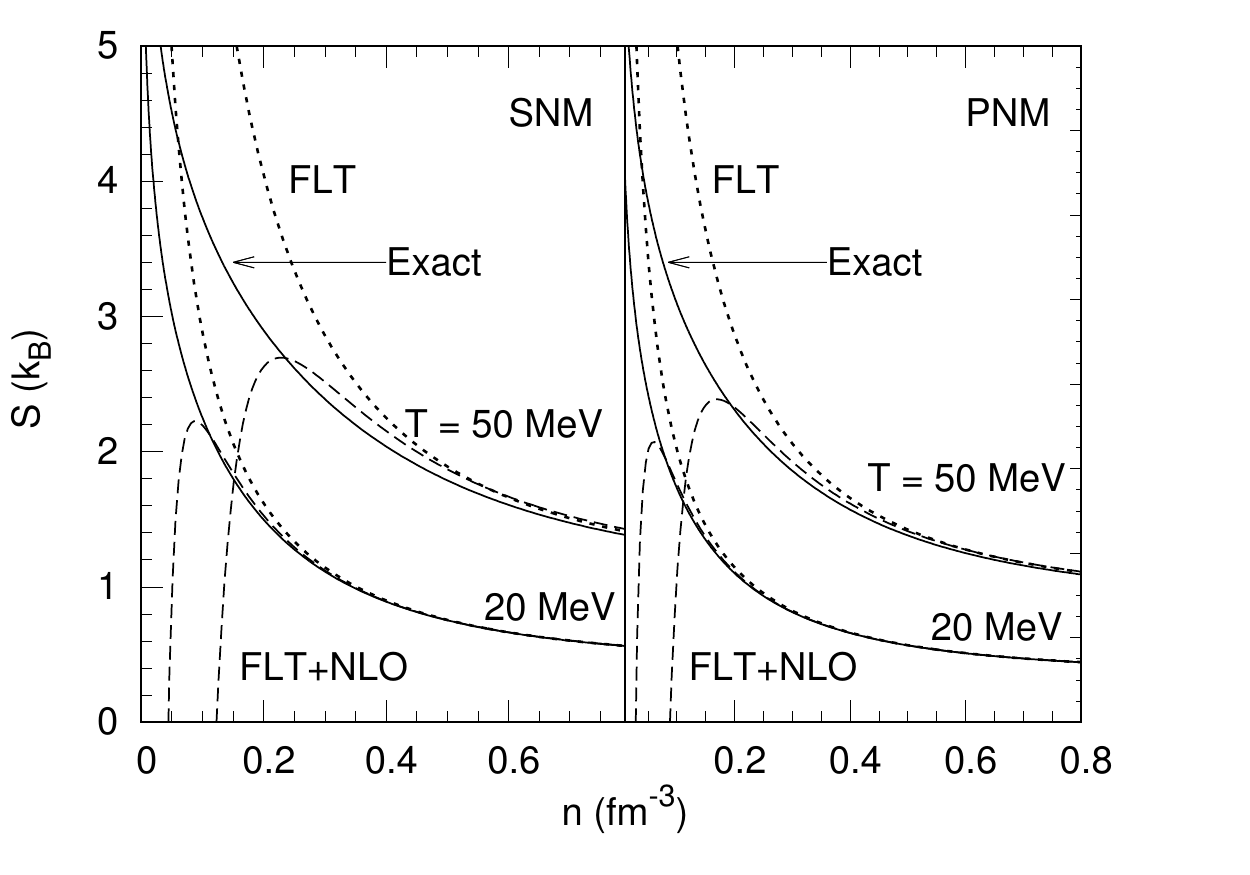}}
\vspace*{-0.2in}
\caption{Entropies per particle $S=s/n$ vs. baryon number density in SNM and PNM. 
Results labeled ``Exact'' are from Ref.~[\refcite{zhang16}]. The leading order Fermi Liquid Theory results are labeled ``FLT'' whereas ``FLT+NLO'' stands for results of 
next-to-leading-order FLT  with $s$ from Eq. (\ref{s-nlo}). Values of temperatures are as indicated in the figure.}
\label{soa}
\end{figure}


In astrophysical phenomena involving supernovae, neutron stars and binary mergers, the entropy per baryon $S$ serves as a gauge to track hydrodynamical evolution  and its 
consequences~[\refcite{BBAL79}].    Figure \ref{soa} shows $S$ vs $n$ in SNM and PNM at $T=20$ and 50 MeV, respectively.  For both SNM and PNM, the NLO corrections substantially 
improve agreement with the exact results for $S$ up to 2. For $T=20$ MeV, the agreement extends to the subnuclear nuclear density of $n=0.1~{\rm fm}^{-3}$ for both SNM and PNM. 
This agreement is encouraging as for $n\simle 0.1~{\rm fm}^{-3}$ and $T \simle 20$ MeV, matter exists in an inhomogeneous  phase consisting of heavy nuclei, light nuclear clusters such 
as $\alpha$ particles, tritons and deuterons, and dripped nucleons (as also leptons and photons) for which a separate treatment is required.  The lesson learned is that up to 
$S=2$, the thermal properties of bulk homogeneous nucleonic matter is adequately described by a knowledge of the $T=0$ spectra of nucleons from which all thermal properties 
can be obtained through the use of FLT carried up to NLO terms.  

\begin{figure}[htb]
\centerline{\includegraphics[width=11cm]{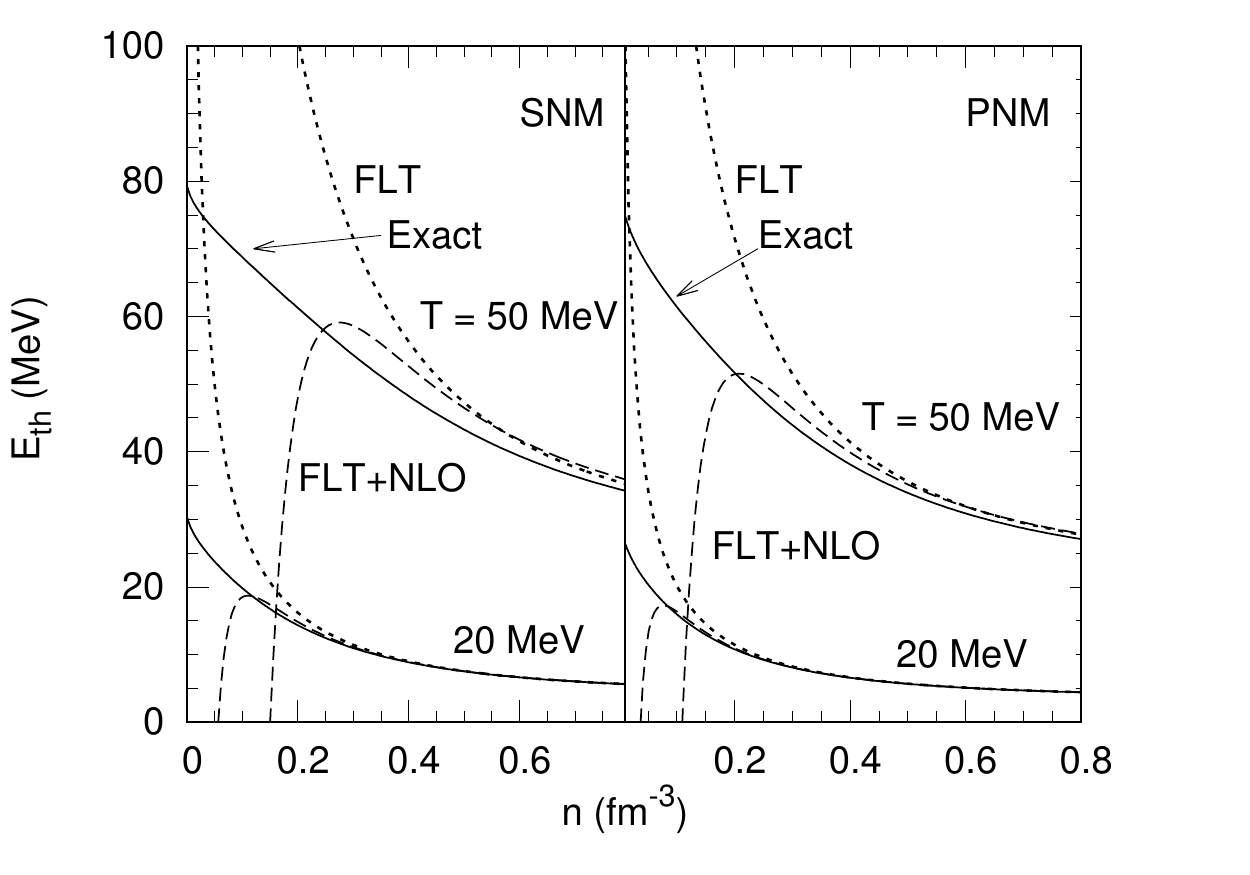}}
\vspace*{-0.25in}
\caption{ Same as Fig.~\ref{soa} but for thermal energies from Eq.~(\ref{Eth-nlo}).}
\label{Eth}
\end{figure}

In Fig.~\ref{Eth}, we show results for the thermal energies. As for $S$, the NLO corrections extend agreement with the exact results down to $n\simeq 0.1~{\rm fm}^{-3}$ in both SNM 
and PNM. The agreement for PNM extends to somewhat lower densities because PNM is more degenerate than SNM at the same $n$. The results at $T=50$ MeV indicate the densities 
for which matter is in the semi- or non-degenerate regions. The substantial improvement offered  by the NLO corrections are, however, noteworthy. 

\begin{figure}[htb]
\centerline{\includegraphics[width=11cm]{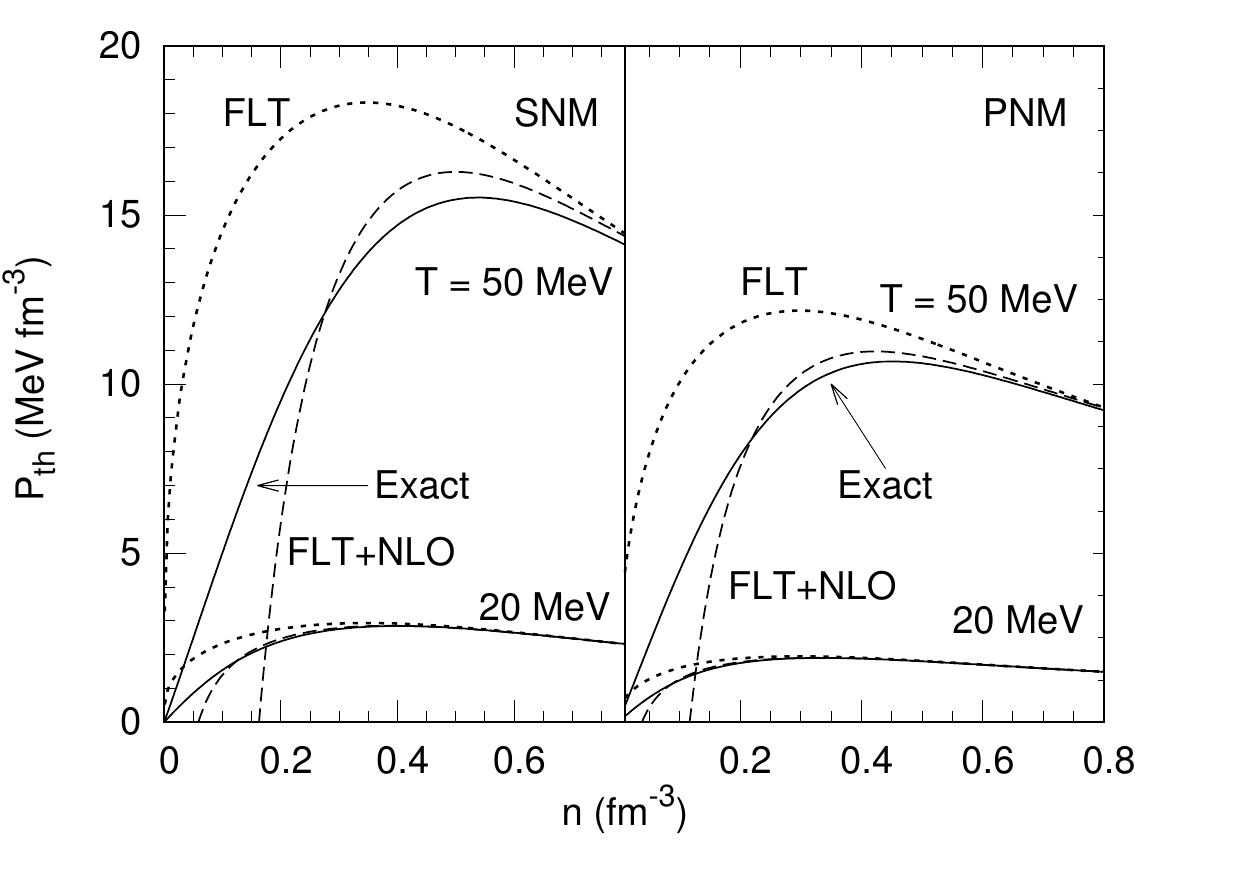}}
\vspace*{-0.25in}
\caption{Same as Fig. \ref{soa}, but for thermal pressure  from Eq.~(\ref{Pth-nlo}).}
\label{pth}
\end{figure}

Figure \ref{pth} contains results for the thermal pressures. As for $S$ and $E_{th}$, agreement of the FLT+NLO results with those of exact numerical calculations extend up 
to $n\simeq 0.1~{\rm fm}^{-3}$ at $T=20$ MeV. The situation with the results at $T=50$ MeV is less satisfactory. The disagreement with the exact results at this temperature is partly owing to the fact 
that $M^*$ begins to acquire a non-negligible temperature dependence as $T$ increases~[\refcite{pabw87}], not considered in the FLT+NLO treatment.  Also at work is the 
fact that the thermodynamic identity ${\mathcal E} + P = Ts + \mu n$ cannot be satisfied even in principle  beyond the Hartree level unless the theory is exactly solved. 

\begin{figure}[htb]
\centerline{\includegraphics[width=11cm]{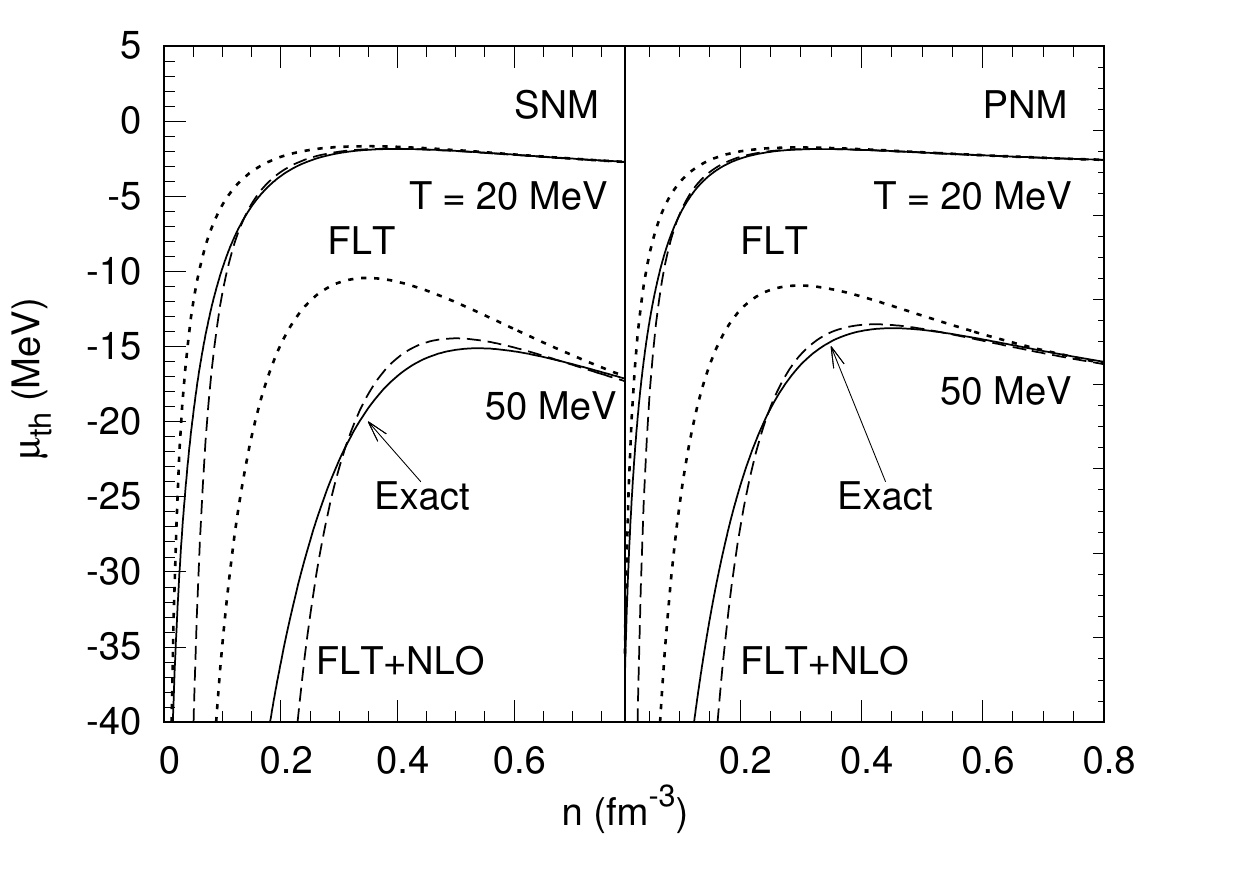}}
\vspace*{-0.25in}
\caption{Same as Fig. \ref{soa}, but for thermal chemical potential  from Eq.~(\ref{muth-nlo}).}
\label{muth}
\end{figure}

The thermal parts of the chemical potentials are shown in Fig.~\ref{muth}. As for  $S,~E_{th}$, and $P_{th}$, the NLO corrections render significant improvement over the FLT 
results down to $n\simeq 0.1~{\rm fm}^{-3}$ for $T=20$ MeV. The agreement of the FLT+NLO results with the exact results is quantitatively better for PNM than for SNM because 
of its higher degeneracy at this temperature at the same density. The NLO improvements at $T=50$ MeV are less striking than at 20 MeV, and suffer from the 
same maladies  as the other thermal variables. Analytic expressions for Fermi integrals being asymptotic expansions, this disagreement  is unavoidable particularly in the 
semi-degenerate region.  A separate treatment as espoused in Ref. [\refcite{pabw87}] is necessary in the non-degenerate region. 

\begin{figure}[htb]
\centerline{\includegraphics[width=11cm]{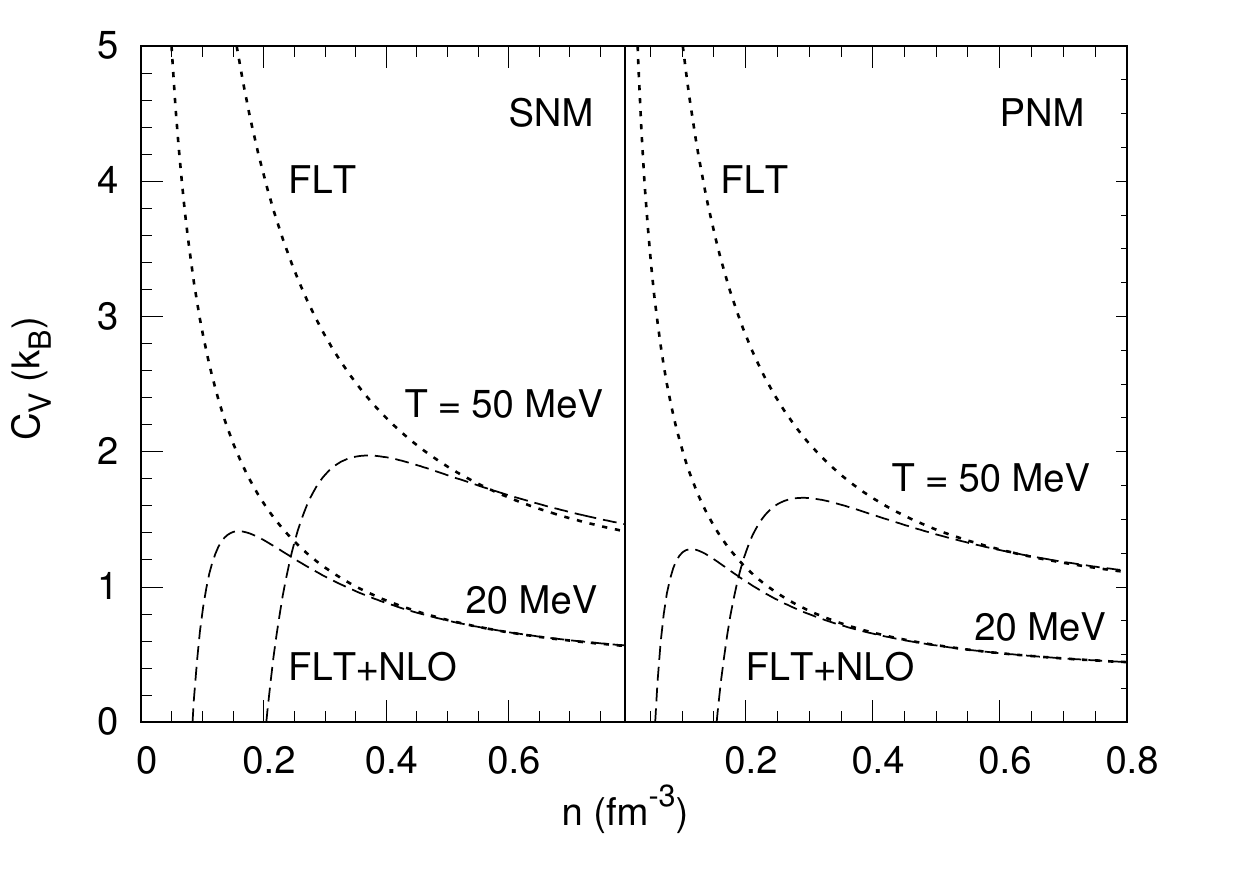}}
\vspace*{-0.25in}
\caption{Same as Fig. \ref{soa}, but for specific heat at constant volume  from Eq.~(\ref{cv-nlo}).}
\label{cv}
\end{figure}

The specific heat at constant volume $C_V$ is shown in Fig.~\ref{cv}. Exact results for $C_V$ were not calculated in Ref.~[\refcite{zhang16}], but we can easily gauge the 
improvement from NLO corrections at near nuclear  densities by recalling that at leading order in FLT, $C_V=S$. The quantity $C_V$ plays a major role in the long-term cooling 
of a neutron star. For example, the time for a star's center to cool by neutrino emission can be estimated by 
\begin{equation}
\Delta t = - \int \frac {nC_V}{\epsilon_\nu}~dT \,,
\label{coolingt}
\end{equation}
where $\epsilon_\nu$ is the neutrino emissivity and $T$ is the temperature. At low temperatures ($T \leq 1$ MeV), however, corrections to $C_V$ arising from Cooper-pairing of 
nucleons must be considered. 

\begin{figure}[htb]
\centerline{\includegraphics[width=11cm]{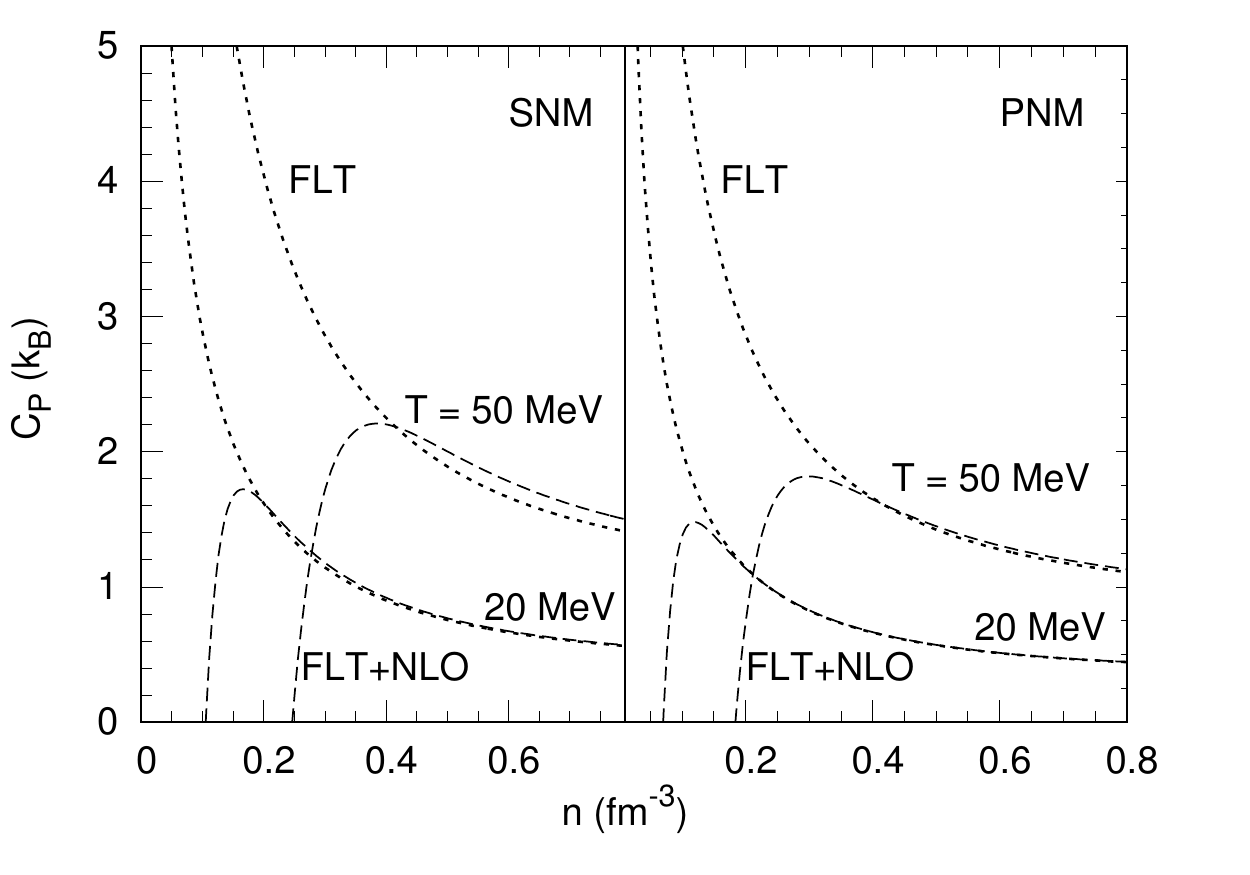}}
\vspace*{-0.25in}
\caption{Same as Fig. \ref{soa}, but for specific heat at constant pressure  from Eq.~(\ref{cp-nlo}).}
\label{cp}
\end{figure}

Figure~\ref{cp} shows results for the specific heat at constant pressure. As for $C_V$, exact numerical results for $C_P$ are not yet available, hence only the FLT and FLT+NLO results are 
shown. At leading order in FLT, $C_P=S$. It is intriguing that at $T=20$ MeV, the NLO corrections do not alter the leading order FLT result down to near nuclear densities in both SNM 
and PNM. While the situation is similar for $T=50$ MeV in PNM for $n\simge 0.35~{\rm fm}^{-3}$, NLO corrections are apparent in SNM.  Exact calculations at $T=50$ MeV would be 
necessary to confirm the extent to which NLO corrections improve the FLT results. A relation similar to Eq.~(\ref{coolingt}) but with $C_V$ replaced by $C_P$ and $\epsilon_\nu$ 
replaced by $\epsilon_{\gamma+\nu}$ is often used in the literature for time estimates in astrophysical phenomena.

\begin{figure}[htb]
\centerline{\includegraphics[width=11cm]{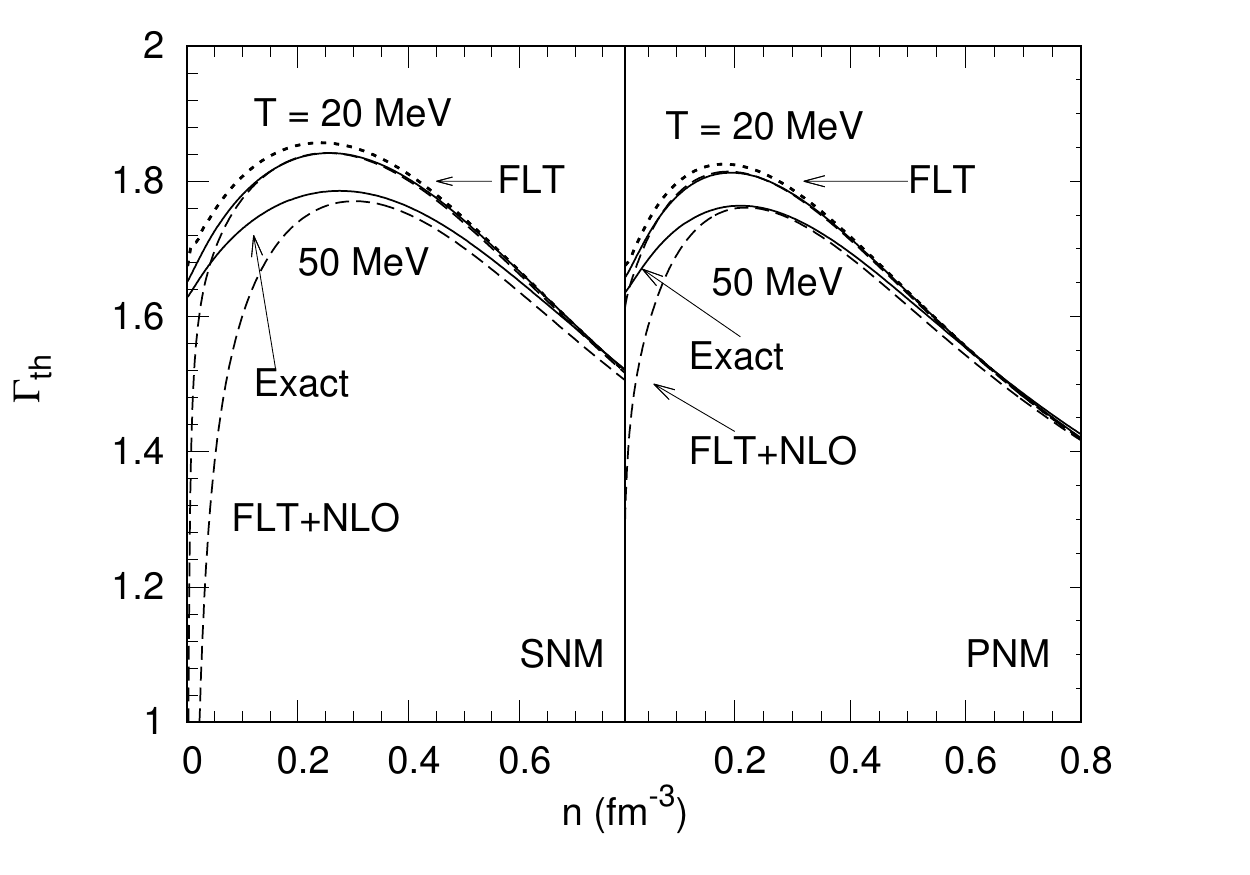}}
\vspace*{-0.25in}
\caption{Same as Fig. \ref{soa}, but for the thermal adiabatic index  from Eq.~(\ref{gth-nlo}).}
\label{gth}
\end{figure}

Results for $\Gamma_{th}$ are shown in Figure~\ref{gth}. At leading order in FLT, ${\displaystyle{\Gamma_{th} = \frac 53 - \frac {n}{m^*} \frac {dm^*}{dn}}}$ and is independent 
of $T$. This feature is borne out by the results (short dashed curves) in both SNM and PNM, the differences between them stemming from differences in the logarithmic  derivatives of 
the Landau effective masses (see Fig.~\ref{msom}). At NLO, $\Gamma_{th}$ acquires a temperature dependence owing to terms proportional to $T^4$ in both $P_{th}$ and $\epsilon_{th}$.  
With increasing $n$, and hence degeneracy, 
the coefficient of the leading ${\cal O}(T^2)$ term in $\Gamma_{th}$ decreases making the NLO corrections to diminish in magnitude.  The agreement of the FLT+NLO results with the exact results extends to sub-nuclear densities at $T=20$ MeV in both SNM and PNM. The $T=50$ MeV results delineate the regions of 
density for which a semi-degenerate analysis is warranted. At very low densities, the exact ${\displaystyle{\Gamma_{th}\rightarrow \frac 53}}$ the value for non-relativistic ideal gases.

\clearpage

From the results shown above in Figs. \ref{soa} through \ref{gth}, it is clear that the lowest density beyond which the FLT+NLO  results reproduce the exact numerical results as a function of increasing temperature steadily increases owing to the semi-degenerate region being encountered. The case of $T=30$ MeV is especially interesting as it happens to be the maximum temperature encountered in core-collapse supernova simulations.  Lacking exact numerical results for MFT+TL calculations at $T=30$ MeV, we performed  exact numerical calculations for MFT using the couplings in Ref.~[\refcite{zhang16}] and compared the ensuing results with those of FLT+NLO (not shown here). For all thermal variables, our findings are: (1) for SNM, very good agreement is found for  $n\simge 0.2~ {\rm fm}^{-3}$, and (2) for PNM, the agreement is very good starting from the nuclear density of $n_0\simge 0.16~ {\rm fm}^{-3}$. 
We expect a similar behavior for MFT+TL because the total single-particle spectra for both SNM and PNM are predominantly composed  of their corresponding MFT parts.  For values of $Y_e$ intermediate to those of SNM and PNM, caution must be exercised in carrying the conclusions above as one or the other nucleonic species may be in the non- or semi-degenerate region.

\section{Summary and Conclusions}
\label{SumCon}

In this work, the next-to-leading order (NLO) extension of Landau's Fermi Liquid Theory (FLT) developed in Ref.~[\refcite{cons15}] was utilized to calculate the thermal properties 
of symmetric nuclear and pure neutron matter (SNM and PNM) for the relativistic model of Ref.~[\refcite{zhang16}]  
 in which two-loop (TL) corrections to mean field theory (MFT) 
were included.  In FLT, the Landau effective mass $m^*$ and its logarithmic derivative with respect to density $n$ suffice to capture the leading order temperature ($T$) effects. The 
NLO corrections, which account for the next-higher-order effects in $T$, require up to second order derivatives of the generalized Landau effective mass function 
${\displaystyle{{\mathcal M}(p)=p\left(\frac {\partial\epsilon}{\partial p} \right)^{-1}}}$, where $\epsilon\equiv\epsilon(n,p)$  is the density and momentum dependent part of the $T=0$ 
single-particle spectrum. The explicit form of  $\epsilon(n,p)$  depends on the specific nature of the $T=0$ many-body calculation performed.  Contrasting examples include models 
with contact or finite range interactions, MFT vs MFT+TL approximations, Bruekner-Hartree-Fock vs Dirac-Brueckner-Hartree-Fock, effective field-theoretical approaches at various 
levels of approximation, etc. For all these cases, the NLO extension enables the calculation of the entropy density and specific heats up to ${\mathcal O}(T/T_F)^3$ whereas the energy 
density, chemical potential and pressure to ${\mathcal O}(T/T_F)^4$ (where $T_F$ is the Fermi temperature) extending the leading order results of FLT.  

Our comparisons of FLT and FLT+NLO results with those of the exact numerical calculations reported in  Ref.~[\refcite{zhang16}] for the relativistic model in which TL effects were 
included reveal that substantial improvements are achieved by the NLO corrections for all thermal variables (entropy, energy, pressure, chemical potential, and specific heats) for 
entropy per baryon $S$ of up to 2. It is noteworthy that the NLO corrections extend agreement with the exact results to sub-nuclear densities of $n \sim 0.1~{\rm fm}^{-3}$ for $T=20$ MeV, 
whereas the FLT results are valid for densities beyond $\sim 0.1~{\rm fm}^{-3}$. Insofar as for $T\simle 20$ MeV and $n\simle 0.1~{\rm fm}^{-3}$, an inhomogeneous  phase consisting of heavy 
nuclei, light nuclear clusters, dripped nucleons, and pasta-like configurations exists which requires a separate treatment, the semi-analytical formulas of the FLT+NLO formalism enables a 
rapid evaluation of thermal effects in bulk homogeneous matter in addition to providing physical insights and checks of time-consuming exact numerical calculations.  

Several areas for further investigation remain including an assessment of non-analytic contributions arising from long-wavelength fluctuations, single particle-hole excitations and, 
collective and pairing correlations close to the Fermi surface~[\refcite{baym91}]. Establishing their roles in astrophysical phenomena needs further work and will be reported elsewhere.

\section{Personal Tributes to Gerry Brown}
\label{Tribs}

\paragraph{Constantinos Constantinou}
When the seminar room of the Nuclear Theory Group at Stony Brook was named in his honor, Gerry was playfully `upset': ``Are they telling me that I should retire?'' 
He certainly had no such plans; it simply wasn't in the stars- collapsing, exploding or otherwise. He would come in (almost) every day full of energy and new ideas 
about problems to solve and just early enough to win his little battle with John Milnor for the $\#2$ YITP parking spot; well... sometimes.

I was fortunate to have been under Gerry's tutelage for about two years. During this time he made it a point that I should learn Fermi Liquid Theory- among many 
other things which would be sprung upon me faster than I could dig out references for. In hindsight, I should have asked Gerry for those; he was much better than 
any search engine in this regard. In our contribution here, we apply an extended version of FLT to a relativistic model beyond the mean-field level. Perhaps Gerry would have liked it.

\paragraph{Madappa Prakash}
Gerry Brown was often fond of saying that although Landau's Fermi Liquid Theory originated in Russia, few Russians used it and it was left for others to exploit Landau's genius. Gerry, along with Kevin Bedell, taught  me this subject which I have used whenever I can to gain physical insights. I recall Gerry struggling to satisfy Landau's forward-scattering sum rule using results from many-body calculations of nuclear matter. He exhorted all who could (particularly, Pandharipande) to provide him with Fermi-liquid parameters and  was disappointed when they could not owing to the inherent difficulties in their many-body methods.  The laments of his toil are recorded in a Physics Reports he wrote with Sven-Olaf B{\"a}ckman and Joni Niskanen [\refcite{brown85}]. 

Our contribution to Gerry's 90$^{\rm th}$ birthday memorial tribute here resulted from my collaboration with Gerry's last graduate student Constantinos Constantinou and my current graduate student Sudhanva Lalit. I like to think that in undertaking the work reported here, I am passing on the lessons Gerry taught me.  He would have been pleased with the extensions to leading-order FLT that was developed by us in Ref. [\refcite{cons15}] and put to good use.  Not one to praise anyone to his or her face, Gerry would have  said  ``What about Landau's forward-scattering sum rule?''.  No doubt that would have annoyed us, but he always wanted to go forward. 

A few ``Gerry-sms'' that I can never forget. ``Don't be a scholar, do things''. This one used to annoy me the most. What annoys me more is that I use it on my own students nowadays!   ``People keep saying they're consistent. But, are they right?'' No arguments there. When warned to be careful, ``I'm never careful! I want to get ahead''.  Bravado there. But, I have 
read his D.Sc. thesis in which he was super-careful.  ``I've no problem using results that I don't understand.''  He selected results of those he trusted. There are more, but for some other time and some other place.

\section{Acknowledgement} 
We are grateful to Xilin Zhang for providing numerical results of the exact two-loop calculations in tabular form. This work was supported  by the U.S. DOE under 
Grant No. DE-GG02-93ER-40756 (for S.L and M.P.)

%

\clearpage

%


\end{document}